# THREE-DIMENSIONAL CONFINEMENT OF VAPOR IN NANOSTRUCTURES FOR SUB-DOPPLER OPTICAL RESOLUTION

**Philippe BALLIN, Elias MOUFAREJ, Isabelle MAURIN,**
**Athanasios LALIOTIS, Daniel BLOCH***

Laboratoire de Physique des Lasers, Université Paris 13, Sorbonne Paris-Cité, France
and
CNRS, UMR 7538, 99 Avenue J-.B. Clément, F-93430 Villetaneuse, France

*We confine a Cs thermal vapor in the interstitial regions of a glass opal. We perfom linear reflection spectroscopy on a cell whose window is covered with a thin film (10 or 20 layers) of ~ 1000 nm (or 400 nm) diameter glass spheres, we observe sub-Doppler structures in the optical spectrum for a large range of oblique incidences. This original feature associated with the inner (3-dimension) confinement of the vapor in the interstitial regions of the opal, evokes a Dicke narrowing. We finally consider possible micron-size references for optical frequency clocks based on weak, hard to saturate, molecular lines.*

* e-mail : daniel.bloch@univ-paris13.fr

Free atom and molecules are the elementary components providing universal frequency references, in the radiofrequency (*r.f.*) and optical domains as well, but the size of the corresponding devices (clocks) is large. The ultimate performances allowed by laser-cooled atoms or ions require huge cooling and trapping volumes. Simpler systems, based upon a thermal vapor, still remain bulky because a low gas density is required to minimize collisions, and also because the free molecules must be kept far away from the surface. For many applications, including communication technologies, there is an urgent need to develop compact frequency references of various levels of accuracy. Pushing conventional vapor technologies up to their minimal size limits, an integrated optical (*r.f.*) clock device as small as a "grain of rice" was demonstrated [1], relying on known macroscopic techniques.

Recently, vapor phase spectroscopy has started to be combined with the realm of micro- and nanostructures [2-13]. Meter-long holey or photonic fibers, sealed with a vapor inside [2-5], have allowed "all-fibered cell" with a two-dimensional (2-D) confinement; aside from the size reduction, a benefit is the possibility of enhanced nonlinear optical effects [2,3]. The pores of a glassy medium have been used as an atomic vapor container, but only the free alkali-metal atoms having escaped outside the porous glass were observed for a 7-30 nm pore size [6]. An interest of such a system is that a specific desorption process allows controlling the atomic density independently of the thermodynamic equilibrium ("LIAD" atomic dispenser). Spectroscopy inside the interstitial regions [7] could be performed for larger pores. Multiple photon scattering increases the effective absorption length, but broad resonances are imposed by the high-density of the molecular gas and by transit time. Until now, the major Doppler broadening, associated to thermal molecular motion, could not be eliminated for these 2-D or 3-D confined vapors, except with a holey fiber, when using a nonlinear scheme of saturated absorption (SA) spectroscopy and a rather large confinement [4,5]. Moreover, tightening the confinement (*e.g.* from a 20 μm diameter core to 10 μm) [4] causes a broadening owing to transit time, in spite of a possible enhancement of the contribution of slow atoms [5].

In the work of our group [8-11], notable spectroscopic advantages, including Doppler-free resonances in nonlinear [8] and linear spectroscopy [9,10], and sensitivity to atom-surface interaction [11], were demonstrated for 1-D confinement (possibly down to <100 nm). This 1-D confinement proves now useful for noticeable demonstrations of collective effects such as Rydberg blockade [12] and cooperative Lamb shift [13], but it uses a bulky technology.

Here, we probe Cs atoms confined in the sub-micronic interstitial regions of an opal made of glass spheres of diameter d ~ 1µm or (alternatively) d ~ 400 nm. Working on the $D_1$ and $D_2$ resonance line doublet of Cs ($\lambda_1$ = 894 nm, $\lambda_2$ = 852 nm), we demonstrate sub-Doppler structures in the linear regime induced by the 3-D confinement of the vapor.

To construct our sealed Cs vapor cells (cylindrical glass tube with a reservoir and two windows, length ~ 4 cm, diameter ~ 4 cm, see fig.1), one of the window is partially coated with a thin opal, self-organized layer-by-layer through a soft chemistry technique of Langmuir-Blodgett deposition [14]. The cohesion of the opal, only resulting from the weak van der Waals contact forces, is further enhanced through annealing. Dispersion in the size of the glass spheres make the opal polycrystalline. An approximate compact arrangement combining *f.c.c.* and *h.c.* structures is expected, with compacity defects increasing with successively deposited layers [14]. Once filled with Cs, the opals become greenish or even dark, a signature of the generation of alkali clusters favored by the trapping of Cs atoms in the interstitial regions of the opal. This effect jeopardized our initial attempts to work with massive opals of small spheres [15]. For thin opals, extra-heating of the window (typically, $T_{window}$ ~180-200 °C for $T_{reservoir} \leq 110°C$) returns the opals (10 or 20 layers) with d ~ 1µm spheres to its initial light milky appearance. For a 10 layers opal cell with d ~ 400 nm spheres, a slight coloring always remains, but consistent results are still obtained. The vapor density, measured by independent spectroscopic experiments, appears unaffected by residual trapping of Cs inside the opal.

Optically, a thin opal is a transparent scattering medium. The transmission for 10 layers is less than 1% when the sphere size (d ~ 1μm) compares with the (near IR) wavelength of Cs (for the opal with d ~ 400 nm spheres, the initial 30-50 % transmission decreases due to the Cs coloring). Remarkably, the "corrugated" nature of the window/opal interface (fig.1) does not prevent the observation of specular reflection from the opal-covered window. Under normal incidence, the reflectivity on this side is comparable to the one (4%) at the flat air/window (input) interface; for oblique incidences, reflectivity becomes polarization-dependent, with a Brewster-like (quasi-null) minimum for TM reflection [16]. The monitoring of reflectivity ("reflection spectroscopy") is the basis of our spectroscopic analysis.

The origin of this reflection is a complex nano-optics problem because scattering occurs here on a dimension comparable to the wavelength. We have developed two approaches, considering either a stratified medium with an effective index governed by the (planar) filling factor of the opal [17], either a rigorous finite element method [18], restricted however to a small (2 or 3) number of layers for numerical convenience. In essence, the "window+opal" system combines the $[(2/3)^{1/2}d]$ periodicity of the opal (assumed to be compact), with a nearly empty gap (thickness d/2) at the window/opal interface (fig.1). The reflectivity, apparently large for a "glass/glass" interface, relates to the size of the gap allowed by the nearly empty contact region between the window and the opal first layer [17,18]. This reflectivity is predicted to depend upon the incidence angle, exhibiting oscillations in its wavelength dependence, and Bragg peaks. These features were experimentally demonstrated with a white source and a spectrum analyzer [17].

To enhance the visibility of narrow spectroscopic contributions, we apply a frequency modulation (FM) to the laser (typically at f ~ 10 kHz, with an amplitude 5-20 MHz), and process the reflected light with a lock-in detector in order to frequency-derive the spectrum [8-11,19]. For d ~ 1μm opals, our findings summarize as follows:

(i) Near normal incidence, (FM) reflection on the opal exhibits a nearly antisymmetric lineshape (fig.2), whose width (~30 MHz) is clearly below the ~ 200 MHz half-Doppler width, although broader than the (5 MHz) natural width of the transition.

(ii) Increasing the incidence angle, the lineshape largely broadens-up.

(iii) For large incidence angles (θ ~ 30-60°, as measured at the external air/window interface), a sub-Doppler structure (≤ 30 MHz) appears, superimposed to the broad Doppler-broadened structure and better seen in TM polarization (fig.3). This narrow structure, whose shape evolves rapidly with the incidence angle, is located at the Doppler-free resonance (determined in an auxiliary SA experiment). The amplitude of this narrow structure remains a fraction of the broad signal after FM detection: in a direct detection, it would hence correspond to a few % of the broad signal. With 2f detection, the broad background nearly vanishes, yielding only the narrow structure (fig.4).

(iv) This sub-Doppler structure appears in the linear regime of incident intensity. We have carefully checked that it does not originate in a residual nonlinear signal (as easily occurring with optical pumping, see *e.g.* [6]), nor in a stray signal as induced by scattering in the opal. Our sensitivity allows us to reach ~10 μW/cm² incident intensity, well below the few mW/cm² Cs saturation intensity in volume [20].

These results are unchanged for cells with 10 and 20 layers, or when investigating various regions of the opal (*i.e.* they are not associated to local defects or orientation in the opal). The Cs density in the interstitial regions appears similar to the one of the free volume, as shown when comparing -after due FM width normalization- the amplitude of the signal with the one at the flat interface (uncoated region of the opal). For the d ~ 400 nm opal, analogous results are obtained in spite of the remaining Cs clusters.

Until now, only two techniques had generated sub-Doppler and linear spectroscopic signals with a vapor, namely : selective reflection (SR) spectroscopy at a flat window/vapor interface [19, 21], and thin cell spectroscopy [9-11]. In both techniques, corresponding to a kind of 1-D confinement (either optical, or mechanical), the Doppler broadening is suppressed only under normal incidence: the dominant contribution comes from atoms flying nearly parallel to the walls, less sensitive to relaxation or transient effects imposed by wall collisions. Here, the sub-Doppler contribution observed at oblique incidences is a feature with no equivalent, which we attribute to the 3-D confinement in the interstitial regions of the opal.

For such an interpretation, we first rule out the contribution of the free vapor/opal interface, analogous to SR at a corrugated interface (after propagation through the entire opal thickness). Indeed, the results are similar for the 20 and 10 layers opal, while the transmission is much weaker for the 20 layers opal. Second, the response of the empty region surrounding the first (half-)layer of glass spheres may evoke a quasi 1-D confinement, instead of a genuine 3-D confinement. This confined region can yield a behavior resembling the one of a vapor nanocell [10], characterized by a sub-Doppler lineshape under normal incidence and a broadening-up with the incidence angle; because of this, we attribute an essential part of the signal observed under small incidence angles to this region. The corrugated interface, like a distribution of incidence angles, may induce a residual broadening, justifying the ~30 MHz width. Conversely, even if some peculiar incidences on the corrugated first half-layer may produce a guided wave (*e.g.* like in plasmonics), one cannot justify that the narrow linear signal originates from this region for oblique incidences spanning on a large range.

Rather, we associate the sub-Doppler contribution at oblique incidences to a probing of atoms confined *inside* the interstices of the opal. Indeed, in reflection spectroscopy, the atomic back-action is as a summing of the local atomic responses (*i.e.* absorption and dispersion), modulated by a complex propagation factor. At a planar interface, the reflected field [21] is governed by a contribution $\int_0^{+\infty} p(z) \exp(i\beta kz) \, dz$, with $\exp(i\beta kz)$ the (oblique) backward propagation factor, and with *p(z)* the complex atomic response at a *z* distance to the surface : *p(z)* includes the spectral dependence over the various atomic velocities and the non local transient response of the moving atoms; here, the phase factor for the incident propagation is included too. For an interface with an opal, an analogous treatment is possible, although more complex. The strong and periodical variations of the number of atoms, governed by the local emptiness factor between spheres, must be included in *p(z)*, as well as the distribution of atomic trajectories in the interstices. Propagation is also more complex: the backward propagation, yielding the reflection, sums up the light coherently *scattered* by the emitters from the periodical interstices; the field driving the atomic polarization *p(z)* is not simply the propagating transmitted field but may have to include the scattered light [18]. Changing the incidence, the wavelength, or even the polarization modifies the spatial distribution of the complex atomic emission, possibly turning a good matching with the periodicity of the opal into a mismatch. A Bragg-type derivation provides a crude estimate [22] of the incidence which yields a constructive contribution from the atoms located in inner interstices. It must be tempered by the initial dephasing associated to the first half-layer of interstices, by the attenuation of the exciting field towards the inner region and by the complex nature of the atomic response (absorption/dispersion). Also, the defects in the opal periodicity [14] strongly affect the visibility of the deepest interstices.

Experimentally, the major argument for the contribution of the inner regions in the narrow spectral contribution is the high sensitivity of the lineshapes to polarization and incidence angle (fig.3). The field propagation being very sensitive to polarization and incidence, a minor change in propagation makes the summing of the atomic response favoring either absorption, or dispersion, and induce sensible changes in the lineshape [23]. A comparison between the $D_1$ and $D_2$ lines confirms this influence of the propagation. In our linear regime, the spectrum of the $D_2$ line, with its hyperfine structure, can be simply predicted from the $D_1$ spectrum with its isolated hyperfine component. A good agreement is indeed found up to those oblique incidences for which a sub-Doppler response appears. Thence, the lineshapes profoundly differ, notably in the mixture of broad and narrow structures. In spite of the tiny difference in wavelengths, propagation has changed critically with the $\lambda/d$ ratio [17].

The inner contribution that we have discussed above is spectroscopically sub-Doppler, and a full interpretation of this point is still pending. Our research was inspired by the Dicke idea [24, 25] that the Doppler broadening disappears when the atomic motion remains confined within a fraction of the wavelength. The Dicke narrowing is found in a variety of situations: for 3-D confinement, it is classically a confinement imposed by buffer gas collisions, which was early observed for thermal gases in *r.f.* transitions [24], and it assumes collisions changing only velocity, with no loss of coherence; conversely, in the Dicke narrowing for a 1D-confined system [9, 10, 25], collisions to the surface induce a complete de-excitation, with all atoms evolving in a transient regime of interaction, and the line-narrowing originates in the (logarithmically) enhanced contribution of (1-D) slow atoms. None of these situations seems to apply clearly to our opal-confined gas. Here, the narrow contribution does not seem to be limited by wall-collisions like in [7], perhaps because of an enhanced contribution of slow atoms [5]. The (3-D) Dicke narrowing with a buffer gas has remained elusive in the optical domain [26], and this was our initial reason to choose opals with $d \sim \lambda$ (~1 µm) spheres [27]. At optical energies, a notable difference between buffer-gas confinement and wall-confinement is that a collision with a perturber atom (buffer gas) induces only a "dephasing" collision because the kinetic energy is too weak to allow inelastic collision; conversely, a genuine energy quenching occurs when an atom collides a hard medium like a glass sphere.

In conclusion, aside from our success in confining the alkali-metal vapor in a way compatible with spectroscopy, our major result is the observation of a sub-Doppler signal in linear spectroscopy, originating from the inner regions of the opal. For applications, a sensitive issue that requires further experimental studies is the stability of the shape of the narrow structure, whose symmetry presently varies with the incidence angle. Alternate to the reflection measurements reported here, a detection of the scattered light could be considered [28], allowing complex or irregular frontiers of the confining nanostructured medium. From our results and as a benefit of the intrinsic linearity of the method combined with the presence of narrow spectral features, one may envision very compact sub-Doppler references based on weak (hardly saturable) molecular lines, possibly applicable to no-drift gas sensors. The probed volume yielding a sub-Doppler response can be conceptually as small as several spheres (*i.e.* several $\lambda^3$), and the irradiation extremely focused. An efficient multiple scattering could also increase the effective optical path inside the resonant vapor, as in [7]. Another extension of our experiments would be to look for an equivalent sub-Doppler contribution in a porous medium, when the average pore size is $< \lambda$. All these self-organized or random confinement systems combine a simpler realization and similar advantages than the macroscopic array of sub-wavelength cells proposed in [29]. Note that as long as the confinement length remains above ~100 nm, the atom-surface interaction ([11] and refs. therein) perturbs only marginally the atomic resonances. Conversely, in a situation far from a gas at thermal equilibrium [30], a Dicke narrowing is reported associated to confinement in small pores, at the expense of an important frequency shift-. In the same manner, the atomic motion tends to wash out the photonic crystal nature of an opal, with its local succession of field enhancement / inhibition. Finally, let us add that in complementary experiments to be reported elsewhere, we observed for long time scales, a light-induced atomic desorption (LIAD) [31] inside the opal, that we attribute to the quasi-porous intimate nature of the glass spheres [32]. This increases the versatility of experiments combining resonant gases and opals.

*Work performed in the frame of the ANR project Mesoscopic gas 08-BLAN-0031, and ECOS-Sud U08E01 (France-Uruguay program). We acknowledge discussions with M. Ducloy and M.-P. Gorza, contributions to early steps of the experimental work by M.-P. Gorza, and to opal measurements by H. Failache (ECOS-SUD U08-E01), discussions on the Dicke narrowing with J.R. Rios Leite (CAPES-COFECUB 740/12), and discussions on the field propagation in the opal with I. Zabkov (PICS#51813 CNRS-Russian Foundation for Basic Physics). The opal was deposited at CRPP-Bordeaux in the S. Ravaine team, and the opal cell built-up by F. Thibout (LKB ENS,) with advices for optical quality by T. Billeton (LPL).*

**FIGURE CAPTIONS**

Fig.1 Schematic of the experiment, with 5 layers of (enlarged) nanospheres. The second and fourth layers are darker to show the compact arrangement.

Fig. 2 Frequency spectrum of the FM reflectivity ($dR/d\omega$ with R the reflectivity, and $\omega$ the frequency) at an opal interface (d = 1μm, 20 layers) under near normal incidence, for the F=4 →F' ={3,4,5} manifold of the Cs line $D_2$ line. The dashed lines are markers for the respective frequencies of the same manifold, as obtained in the reference saturated absorption spectrum. The vertical scale is normalized to the non resonant reflectivity. FM excursion: 20 MHz; Cs temperature: 90°C

Fig. 3 Same as fig.2, for the $D_1$ resonance line, with the dashed line for the center of the equivalent saturated absorption spectrum. External incidence angle θ (see fig. 1) as indicated; FM excursion: 15 MHz, Cs temperature: ~110°C. (a): TE polarization; (b): TM polarization.

Fig. 4 Comparison between the first ($dR/d\omega$) and second ($d^2R/d\omega^2$) harmonic detection for the frequency spectrum of reflection on the 20 layers d =1 μm opal ($D_1$ line). TM polarization; θ = 40°, FM excursion: 20 MHz; Cs temperature: 120 °C; intensity ~ 50 μW/cm².

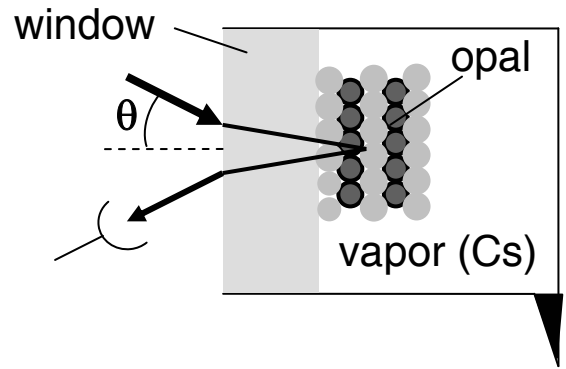

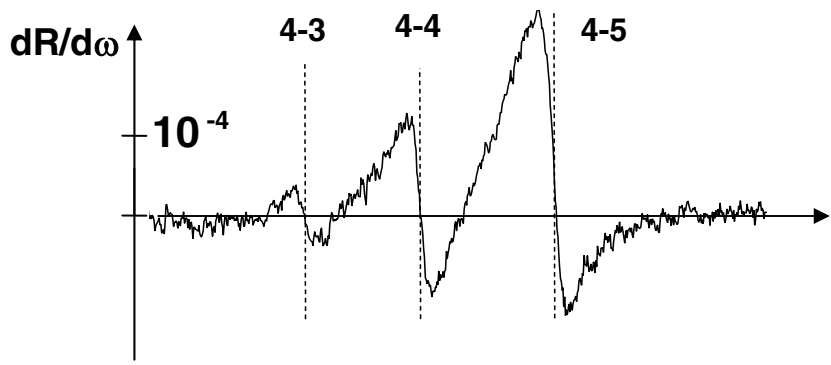

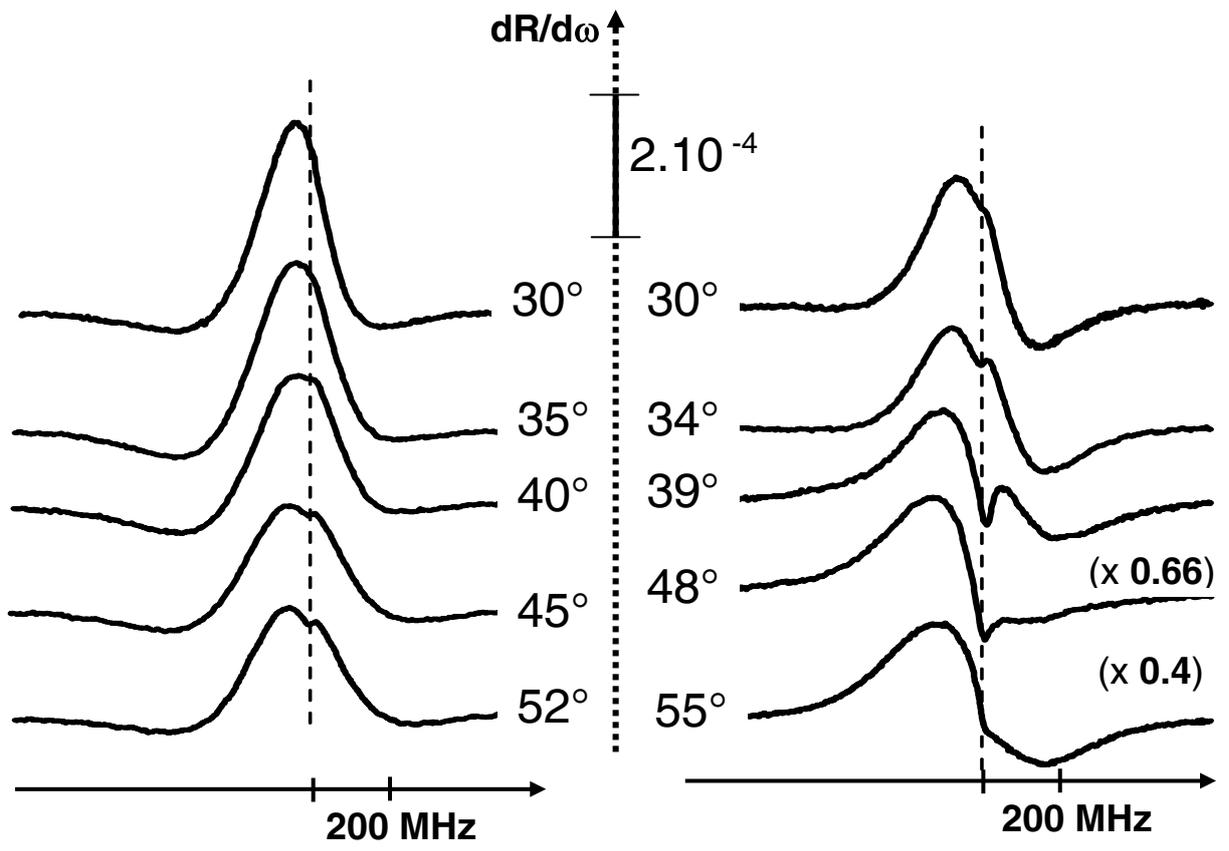

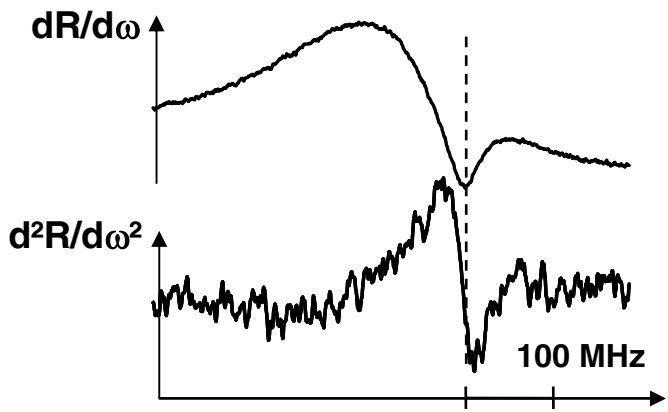